\documentclass[12pt,a4paper]{article}
\pdfoutput=1
\usepackage{a4wide}
\usepackage[centertags]{amsmath}
\usepackage{amssymb}
\usepackage{cite}
\usepackage{ulem}
\usepackage{float}
\usepackage[pdftex]{graphicx,color} 
\allowdisplaybreaks
\begin{document}

\thispagestyle{empty}

\begin{center}
{\Large  \bf Nearly $\mathbf{AdS_2}$ Sugra and the Super-Schwarzian} 
\end{center}

\vspace*{1cm}

\centerline{Stefan F{\"o}rste and Iris Golla}
\vspace{1cm}

\begin{center}{\it
Bethe Center for Theoretical Physics\\
{\footnotesize and}\\
Physikalisches Institut der Universit\"at Bonn,\\
Nussallee 12, 53115 Bonn, Germany}
\end{center}

\vspace*{1cm}

\centerline{\bf Abstract}
\vskip .3cm
In nearly $AdS_2$ gravity the Einstein-Hilbert term is supplemented by
the Jackiw-Teitelboim action. Integrating out the bulk metric gives
rise to the Schwarzian action for the boundary curve. In the present
note, we show how the extension to supergravity leads to the
super-Schwarzian action for the superspace boundary.

\vskip .3cm

\newpage
\section{Introduction}

The theory of gravity simplifies in lower dimensions and thus lower
dimensional gravity could serve as a playground for addressing
difficult problems such as e.g.\ in black hole physics. However, the
simplification might go too far away from a semi realistic
situation. In two dimensions the Einstein tensor vanishes identically,
gravity is not dynamical. To capture some non trivial behaviour often
an additional scalar field (dilaton) is coupled to the
Einstein-Hilbert term and thought of as arising in compactifications
of higher dimensional theories of gravity. (For a review on two
dimensional models see \cite{Grumiller:2002nm}.) A particularly simple
model is provided by the Jackiw-Teitelboim action
\cite{Jackiw:1984je,Teitelboim:1983ux}. Variation with respect to the
dilaton forces the metric to be of constant curvature (whose value is
an input parameter). In the Einstein equation (obtained by considering
metric variations) the Einstein tensor is replaced by the dilaton's
energy momentum tensor. Jackiw-Teitelboim gravity on $AdS_2$ spaces was
recently analysed in \cite{Almheiri:2014cka} in the context of holography. 
In
\cite{Turiaci:2016cvo,Jensen:2016pah,Maldacena:2016upp,Cvetic:2016eiv}
it was shown how the Schwarzian appears as  
an effective Lagrangian for a UV regulator brane on the $AdS_2$
side. This is interesting because the Schwarzian also arises as a
Lagrangian of a Goldstone boson associated to broken reparameterisation
invariance of the SYK model
\cite{Sachdev:1992fk,Kitaev,Sachdev:2010um,Sachdev:2015efa,Polchinski:2016xgd,Jevicki:2016ito,Maldacena:2016hyu}
and alternatives without disorder
\cite{Gurau:2016lzk,Witten:2016iux,Klebanov:2016xxf}. Hence, nearly
$AdS_2$ gravity (or Jackiw-Teitelboim gravity) is related to these
models by holographic duality. Further, black hole physics has been
connected via the SYK model or its alternatives to random matrix
theory
\cite{Garcia-Garcia:2016mno,Cotler:2016fpe,Krishnan:2016bvg,Krishnan:2017ztz}.

In \cite{Fu:2016vas} supersymmetric
versions of the SYK model were shown to lead to super-Schwarzian
Lagrangians. Again there are alternatives without disorder
\cite{Peng:2016mxj} as well as connections to random matrix models
\cite{Li:2017hdt}.  Path integrals with the Schwarzian as well as
the super-Schwarzian action have been recently evaluated
in\cite{Stanford:2017thb}.

An obvious expectation is that the 
super-Schwarzian arises as an effective Lagrangian for a UV regulator
brane in nearly $AdS_2$ supergravity. In the present note, we fill in
the details of this expectation for the case of minimal
supersymmetry. In the next section, we will review the part of
\cite{Turiaci:2016cvo,Jensen:2016pah,Maldacena:2016upp,Cvetic:2016eiv}
which will be modified to include supersymmetry. 
\section{Recap: Schwarzian from Nearly 
$ \mathbf{AdS_2}$  Gravity}  
Here, we repeat the argumentation of \cite{Maldacena:2016upp} (see also
\cite{Turiaci:2016cvo,Jensen:2016pah,Cvetic:2016eiv}) which
will be supersymmetrised in the next section. Our starting point is
the gravity action whose relevant part is (Euclidean signature)
\begin{equation}\label{eq:JT}
S= -\frac{1}{16\pi G} \left[ \int_M d^2x \sqrt{g} \phi \left( R +
    2\right) + 2 \int_{\partial M} \!du\,\sqrt{h}\, \phi K\right] ,
\end{equation}
where $\phi$ is the dilaton and the last term is an adaption of the
Gibbons-Hawking-York boundary term ensuring that there are no boundary
contributions to $\delta S$ once we impose Dirichlet conditions on
variations. The extrinsic curvature, $K$, will be discussed
later. Variation w.r.t.\ $\phi$ yields the constraint $R=-2$
which is solved by the Euclidean $AdS_2$ metric
\begin{equation}\label{eq:metric}
ds^2 =\frac{dt^2 + dy^2}{y^2} .
\end{equation}
Variation w.r.t.\ the metric forces the energy momentum tensor of the
dilaton to vanish leading to the general solution
\begin{equation}\label{eq:dilsol}
\phi = \frac{\alpha + \gamma t + \delta \left(t^2 + y^2\right)}{y} .
\end{equation} 
The other ingredient is introducing a UV brane, i.e.\ a curve which
approaches the boundary at $y=0$ once a small parameter, $\epsilon$,
is taken to zero. The authors of \cite{Maldacena:2016upp} impose the
crucial condition that the proper length of the boundary curve is
constant for finite $\epsilon$. Parameterising the boundary as
$\left( t(u), y(u)\right)$, this leads to
\begin{equation}\label{eq:bc}
\frac{1}{\epsilon^2} = \frac{ {t^\prime}^2 + {y^\prime}^2}{y^2} .
\end{equation}
For later use, we include the first subleading (as $\epsilon \to 0$)
contribution into the solution 
\begin{equation}\label{eq:bcurve}
t= t(u) \,\,\, ,\,\,\, y = \epsilon t^\prime\left( 1 +
  \frac{\epsilon^2}{2} \left(
    \frac{t^{\prime\prime}}{t^\prime}\right)^2\right) +\ldots .
\end{equation}
The easiest argument for the Schwarzian, which just needs the leading
order solution, is to start with a symmetry based guess 
\begin{equation} \label{eq:schw}
S[t(u)] \sim \int du\, \phi_r ( u)\, \text{Sch}\left(
    t,u\right), \end{equation}
where
\begin{equation}
\text{Sch}(t,u) = \frac{t^{\prime\prime\prime}}{t^\prime} - \frac{3}{2}\left(
  \frac{t^{\prime\prime}}{t^\prime}\right)^2 
\end{equation}
denotes the Schwarzian.
Here, $\phi_r (u)$ is a scalar which is related to the dilaton as will
be discussed shortly. We see that the action (\ref{eq:schw}) is
invariant only under an $\text{SL}\left(2 ,{\mathbb R}\right)$
subgroup of reparameterisations.
Variation w.r.t.\ $t(u)$ leads to the equation of motion
\begin{equation}\label{eq:eom}
\left\{ \frac{1}{t^\prime} \left[ \frac{\left(t^\prime
      \phi_r\right)^\prime}{t^\prime} \right]^\prime\right\}^\prime = 0
.
\end{equation}
Taking $\phi_r$ to be the renormalised (i.e.\ multiplied by $\epsilon$)
boundary value of the dilaton (to leading order in $1/\epsilon$)
\begin{equation}
\phi_r\left( u\right) = \frac{\alpha + \gamma t(u) + \delta
  t(u)^2}{t^\prime(u)} 
\end{equation}
shows that (\ref{eq:eom}) is solved for any $t(u)$ which is consistent
with our earlier considerations not leading to further conditions on
$t(u)$. 

A more direct way to obtain the Schwarzian is to evaluate the 2d
action (\ref{eq:JT}) at the metric (\ref{eq:metric}) while cutting off
the integration at the UV brane (\ref{eq:bcurve}). This yields
\begin{equation}
-\frac{1}{8\pi G \epsilon^2}\int du\, \phi_r (u) K.
\end{equation}
The result for $K$ is given in \cite{Maldacena:2016upp} as
\begin{equation}\label{eq:kres}
K= 1 + \epsilon^2 \text{Sch}(t,u) ,
\end{equation}
and we see that after losing an additive divergent contribution we get
indeed (\ref{eq:schw}) now with the prefactor fixed. Since we are
going to supersymmetrise also this calculation let us fill in some
details on the computation of $K$. The extrinsic curvature is defined as
\begin{equation}\label{eq:ecurvdef}
K = g^{\mu\nu}\nabla_\mu n_\nu ,
\end{equation}
where $\nabla$ denotes the covariant derivative and $n_\mu$ is the
normal vectorfield of the UV brane. Its tangent vector is computed by
taking the partial $u$ derivatives of its coordinates
\begin{equation}
T^\mu = \left( t^\prime , \epsilon t^{\prime\prime}\left[ 1 +
      \epsilon^2  \left( \frac{t^{\prime\prime\prime}}{t^\prime} -
        \frac{1}{2} \left(
          \frac{t^{\prime\prime}}{t^\prime}\right)^2\right)\right]
  \right) ,
\end{equation}
where we used (\ref{eq:bcurve}) and stopped writing dots indicating the
existence of higher order terms in $\epsilon$. The normal vector is
defined by the two conditions
\begin{equation}\label{eq:nordef}
T^\mu n_\mu = 0 \,\,\, ,\,\,\, n^\mu n_\mu = 1 .
\end{equation}
Explicitly one finds
\begin{equation}
n_\mu = \left( \frac{t^{\prime\prime}}{{t^\prime}^2} \left[ 1 +
  \epsilon^2 \text{Sch}(t,u)\right] , -\frac{1}{\epsilon
  t^\prime}\left[ 1 - \epsilon^2 \left(
    \frac{t^{\prime\prime}}{t^\prime}\right)^2\right]\right) .
\end{equation}
Because of the normalisation condition in (\ref{eq:nordef}) one can
replace the metric in (\ref{eq:ecurvdef}) by a projector on directions
orthogonal to $n$, i.e.\ on tangent directions. In formulas this is
expressed as
\begin{equation}\label{eq:excu}
K = \left( g^{\mu\nu} -n^\mu n^\nu\right) \nabla_\mu n_\nu = \frac{T^\mu
  T^\nu} {T^2}\nabla_\mu n_\nu = \frac{T^\nu}{T^2}\nabla_T n_\nu ,
\end{equation}
with the directional covariant derivative
\begin{equation}
\nabla_T n_\mu =\frac{\partial n_\mu}{\partial u} - \Gamma^{\rho}
_{\mu\nu}n_\rho T^\nu .
\end{equation}
Here, the normalisation of the tangent vector is fixed by
$T^\mu\partial_\mu =\partial_u$.  
It is now an easy exercise to reproduce
(\ref{eq:kres}). Note, that subleading contributions matter in the
$\Gamma^y_{tt} n_y {T^t}^2$ contribution. The form of $K$ in
(\ref{eq:excu}) will lead us to a supersymmetric version of $K$ in the
next section. 

\section{Super-Schwarzian from Nearly 
$\mathbf{AdS_2}$ Supergravity}    

The supersymmetric version of Jackiw-Teitelboim gravity has been
formulated in \cite{Chamseddine:1991fg}. Conveniently one uses two
dimensional superspace spanned by the coordinates $z, \bar{z}, \theta,
\bar{\theta} $, where $z, \bar{z}$, ($\theta , \bar{\theta}$) values
are complex (Grassmann) numbers related by complex conjugation. Fields
are promoted to superfields depending on superspace coordinates. The
supersymmetric version of (\ref{eq:JT}) reads \cite{Chamseddine:1991fg}
\begin{equation}\label{eq:Nsugra}
S = -\frac{1}{16\pi G} \left[ \text{i}\int d^2z d^2\theta E \Phi \left( R_{+-}
    -2\right) + 2 \int_{\partial M} du d\vartheta \Phi K \right] ,
\end{equation}
where $E$ is the superdeterminant of the vielbein in superspace,
$\Phi$ is the dilaton superfield, $R_{+-}$ is a superfield containing
the usual scalar curvature in the coefficient at the
$\theta\bar{\theta}$ term when Taylor expanded in Grassmann
coordinates, and finally the last term is a supersymmetric version of
the Gibbons-Hawking-York term where $u$ and $\vartheta$ are
coordinates on one dimensional superspace. More details will be
discussed later. Since the author 
of \cite{Chamseddine:1991fg} had closed string worldsheets in mind
this last term is not discussed there.
The boundary curve is now described by two dimensional superspace
coordinates being functions of $u$ and $\vartheta$. We replace condition
(\ref{eq:bc}) by its natural supersymmetrisation
\begin{equation}\label{eq:subc}
\frac{ du^2 + 2 \vartheta d\vartheta du}{4\epsilon^2} = dz^\xi E_\xi ^1
dz^\pi E_\pi ^{\bar{1}} ,
\end{equation}
where Greek (Einstein) indices run over two dimensional superspace
coordinates, whereas $1$ and $\bar{1}$ are flat space Euclidean
indices. The $E$'s denote vielbein components and on the left
hand side the corresponding expression for one dimensional flat
superspace has been divided by the constant $4 \epsilon^2$ (the factor
four is there for later convenience). 
We fix the superconformal gauge
\begin{align}
E_+ & = \text{e}^{-\frac{\Sigma}{2}}D_\theta \equiv
\text{e}^{-\frac{\Sigma}{2}} \left( \frac{\partial}{\partial \theta }
  + \theta \frac{\partial}{\partial z}\right) \,\,\, ,\,\,\,
E_1  = \frac{1}{2}\left\{ E_+, E_+\right\} , \\
E_{-}  & = \text{e}^{-\frac{\Sigma}{2}}D_{\bar{\theta}} \equiv
\text{e}^{-\frac{\Sigma}{2}} \left( \frac{\partial}{\partial \bar{\theta} }
  + \bar{\theta} \frac{\partial}{\partial \bar{z}}\right)  \,\,\, ,\,\,\, 
E_{\bar{1}} =\frac{1}{2} \left\{ E_-, E_-\right\} ,
\end{align}
where braces denote the anti-commutator. In this gauge the supercurvature
is \cite{Chamseddine:1991fg},
\begin{equation}
R_{+-} = -2 \text{i}\text{e}^{-\Sigma} D_\theta D_{\bar{\theta}} \Sigma .
\end{equation}
The constraint $R_{+-} =2$ is solved by the superconformal factor
\begin{equation}
\text{e}^\Sigma = \frac{1}{2\text{Im} z}\left( 1
  -\frac{\text{i}\theta\bar{\theta}}{2 \text{Im} z}\right) . 
\end{equation}
The supervielbein is a four by four matrix with a diagonal two by two
block structure. The holomorphic block is
\begin{equation}
\left( \begin{array}{cc} E_+ ^\theta & E_+ ^z \\
E_1 ^\theta & E_1 ^z \end{array}\right) = \left(\begin{array}{cc}
\text{e}^{-\frac{\Sigma}{2}} &  \theta \text{e}^{-\frac{\Sigma}{2}} \\
\text{e}^{-\frac{\Sigma}{2}} D_\theta \text{e}^{-\frac{\Sigma}{2}} &
\text{e}^{-\Sigma} - \theta  \text{e}^{-\frac{\Sigma}{2}}D_\theta
\text{e}^{-\frac{\Sigma}{2}} \end{array}\right) .
\end{equation}
Its inverse is easily computed to
\begin{equation}
\left( \begin{array}{cc} E^+ _\theta &  E^1 _\theta   \\
E_z ^+ & E_z ^1\end{array}\right) = \left( \begin{array}{cc}
\text{e}^{\frac{\Sigma}{2}} + \theta \text{e}^\Sigma D_\theta
\text{e}^{-\frac{\Sigma}{2}} & -\theta \text{e}^\Sigma \\
-\text{e}^\Sigma D_\theta \text{e}^{-\frac{\Sigma}{2}} & \text{e}^\Sigma 
\end{array}\right) .
\end{equation}
For the anti-holomorphic block corresponding expressions
hold. Plugging all this into (\ref{eq:subc}) yields
\begin{align}
\frac{du^2 + 2 \vartheta d\vartheta du}{4\epsilon^2} = &
  \,\,\,\text{e}^{2\Sigma}\left\{\left|\frac{\partial z}{\partial u}  
    +\theta \frac{\partial \theta}{\partial u}\right| ^2 du^2 + \left[
    \left( \theta \frac{\partial \theta}{\partial \vartheta}-
      \frac{\partial z}{\partial \vartheta} \right) \left(
      \frac{\partial \bar{z}}{\partial u} + \bar{\theta}\frac{\partial
      \bar{\theta}}{\partial u}\right)\right.\right. \nonumber \\
& \left.\left. + \left( \frac{ \partial
    z}{\partial u} + \theta\frac{\partial 
      \theta}{\partial u}\right)\left( \bar{\theta} \frac{\partial
      \bar{\theta}}{\partial \vartheta}- \frac{\partial \bar{z}}{\partial
      \vartheta} \right)\right] d\vartheta d u \right\} .\label{eq:uvbrane}
\end{align}
The ratio between the two components on the RHS is
compatible with the LHS if
\begin{equation}\label{eq:intcond}
Dz = \theta D \theta \,\,\, ,\,\,\, D\bar{z} =
\bar{\theta}D\bar{\theta} ,
\end{equation}
where $D$ denotes the one dimensional superderivative
\begin{equation}
D \equiv \frac{\partial}{\partial \vartheta} + \vartheta
\frac{\partial}{\partial u} .
\label{eq:oneD}
\end{equation}
After imposing (\ref{eq:intcond}) the first iterated solution to
(\ref{eq:uvbrane}) reads
\begin{equation}
\theta =\bar{\theta} = \xi \,\,\, ,\,\,\, z = t + \text{i} \epsilon
\left( D\xi\right)^2 ,
\label{eq:lead}
\end{equation}
where $t$ and $\xi$ are functions of $u$ and $\vartheta$ satisfying
\begin{equation}\label{eq:scft}
Dt = \xi D\xi .
\end{equation}
Eq.\ (\ref{eq:scft}) can be solved in terms of a commuting function
$f$ and an anti-commuting function $\eta$ \cite{Fu:2016vas} ,
\begin{equation}\label{eq:scftsol}
t = f\left( u + \vartheta \eta\right) \,\,\, ,\,\,\, \xi =
\sqrt{f^\prime (u)}\left[ \eta (u) +\vartheta\left( 1
    +\frac{\eta(u)\eta^\prime (u)}{2}\right)\right] .
\end{equation}
A natural guess for an effective action of the functions $f$ and
$\eta$ is
\begin{equation} \label{eq:guess}
S \sim \int dud\vartheta \Phi_r\!\left(u,\vartheta\right) S\left[ t, \xi ;
u,\vartheta\right] ,
\end{equation}
where we adopted the notation of \cite{Fu:2016vas} for the
super-Schwarzian, i.e.\
\begin{equation}
S\left[ t, \xi ; u,\vartheta\right] = \frac{D^4\xi}{D\xi} - 2
\frac{D^3 \xi D^2 \xi}{\left( D \xi\right)^2} ,
\label{eq:supersch}
\end{equation}
with $\xi$ given by (\ref{eq:scftsol}). $\Phi_r$ is a renormalised
version of the dilaton superfield evaluated at the boundary. Before
determining it by the sugra equations let us derive equations of
motion from (\ref{eq:guess}). To this end, it is useful to decompose
\begin{equation}
\Phi_r\left( u ,\vartheta\right) = \phi_r (u) + \vartheta \lambda_r (u) .
\end{equation}
Eq.\ (\ref{eq:guess}) takes the form
\begin{equation}
S \sim  \int du \left( \phi_r S_b\left( t, \xi; u, \vartheta\right)
  +\lambda_r  S_f\left( t, \xi; u, \vartheta\right)\right) ,
\end{equation}
where the super-Schwarzian components are given by
\begin{align}
S_b & = \frac{1}{2} \text{Sch}\left( f,u\right) \left( 1 -\eta
      \eta^\prime\right) +\frac{ 3 \eta^\prime\eta^{\prime\prime}}{2}
      +\frac{\eta\eta^{\prime\prime\prime}}{2} , \\
S_f & = \text{Sch}\left( f,u\right)\frac{\eta}{2} +\eta^{\prime\prime}
      +\frac{\eta\eta^\prime\eta^{\prime\prime}}{2} . 
\end{align}
The equations of motion are obtained by variations of $f$ and $\eta$
\begin{align}
0 & = \left\{ \frac{1}{f^\prime} \left[ \frac{\left(f^\prime
      \left[\phi_r\left( 1 -\eta\eta^\prime\right)
    +\lambda_r\eta\right]\right)^\prime}{f^\prime}
    \right]^\prime\right\}^\prime   ,\label{eq:sueq1}\\
0 & = \phi_r \text{Sch}\left( f ,u\right)\eta^\prime +\left( \phi_r
    \text{Sch}\left( 
    f,u\right) \eta\right)^\prime + 3 \left( \phi_r
    \eta^{\prime\prime}\right)^\prime + 3 \left(
    \phi_r\eta^\prime\right)^{\prime\prime} -\phi_r
    \eta^{\prime\prime\prime} -\left(\phi_r
    \eta\right)^{\prime\prime\prime} \nonumber \\
& \,\,\,\,\,\, +\lambda_r \text{Sch}\left( f ,u\right)
    +2\lambda_r^{\prime\prime} +\lambda_r\eta^\prime\eta^{\prime\prime}
    +\left(\lambda_r \eta\eta^{\prime\prime}\right)^\prime  +\left(
    \lambda_r \eta \eta^\prime\right)^{\prime\prime} .\label{eq:sueq2}
\end{align}
It remains to determine the boundary values of the dilaton. Using the
details given in \cite{Chamseddine:1991fg} it is easy to see that 
(\ref{eq:dilsol}) is again a solution to the bulk equations up to
terms subleading as the boundary is approached (e.g.\ terms containing
$\theta\bar{\theta}$). The renormalisation prescription is to multiply
with $\epsilon$ yielding
\begin{equation} 
\Phi_r = \frac{ \alpha + \gamma t + \delta t^2}{\left( D\xi\right)^2}
.
\end{equation}
Plugging in (\ref{eq:scftsol}) leads to
\begin{equation}
\phi_r = \frac{\alpha + \gamma f +\delta f^2}{f^\prime}\left( 1 - \eta
  \eta^\prime\right) \,\,\, ,\,\,\, \lambda_r = \gamma \eta + 2 \delta
\eta f -\left( \frac{f^{\prime\prime}\eta}{{f^\prime}^2} + 2
\frac{\eta^\prime}{f^\prime} \right) \left( \alpha + \gamma f + \delta
f^2\right) .
\end{equation}
Consistency can be established by checking that (\ref{eq:sueq1}) and
(\ref{eq:sueq2}) are satisfied which is indeed the case as can be seen
by a straightforward, though tedious, calculation.  

To obtain the super-Schwarzian in a more direct way we have to specify
the last (boundary) term in the action (\ref{eq:Nsugra}). We do so by
covariantising with respect to superspace. To this end, we consider
expression (\ref{eq:excu}) where we replace $\partial _u$ by $D$ (see
(\ref{eq:oneD})) such that its transformation under
super-reparameterisations cancels the Berezinian of the super-line
measure \cite{Fu:2016vas}. Further, we need to change from Einstein
indices to Lorentz indices by means of the two dimensional
supervielbein and hence replace the pulled back Christoffel symbols
by their spin connection equivalent. This leads to
\begin{equation}
K =  \frac{T^A D_T n_A}{T^AT_A} ,
\label{eq:superex}
\end{equation}
with  $A\in \left\{ 1, \bar{1}\right\}$ and
\begin{equation}
D_T n_A   = D n_A + \frac{\partial z^\xi}{\partial\vartheta}\Omega_\xi +
\vartheta\frac{\partial z^\xi}{\partial u}\Omega_\xi ,
\end{equation}
where $\xi$ is a two dimensional (curved) super-space index ( $\xi \in
\left\{ \theta, z,\bar{\theta},\bar{z}\right\}$). Further, we should
replace\footnote{This looks a bit ad hoc but can be motivated as
  follows. Condition 
(\ref{eq:uvbrane}) implies an induced one dimensional supervielbein $e_+ =
\sqrt{2\epsilon} D$ in superconformal gauge. Its super-determinant is
$1/\sqrt{2\epsilon}$. With that measure we should replace $D$ by
$e_+$ which cancels the superdeterminant from the measure. 
Another power of $1/\epsilon$ appears after changing the tangent vector
normalisation to one (see (\ref{eq:tangentvec})). (In the bosonic case
the effects of replacing 
$\partial_u$ by $e_u$ and changing the normalisation of the tangent
vector cancel.) 
Together with $\Phi =
\Phi_r/ \epsilon$ one obtains the factor introduced in
(\ref{eq:epspow}). We will not make use of the one dimensional
vielbein in the rest of the paper.}  
\begin{equation}
du d\vartheta \Phi \rightarrow \frac{1}{2\epsilon^2} du d\vartheta
\Phi_r .
\label{eq:epspow}
\end{equation}
For the computation of the surface term it turns out that first order
corrections in $\epsilon$ are sufficient. However, in addition to
(\ref{eq:lead}) we also need the linear correction to $\theta$,
\begin{equation}
\theta = \xi + \text{i}\epsilon\rho +\ldots\, .
\label{eq:tansatz}
\end{equation}
It is important to notice that this leads to a correction of $\text{Im} z$ at
linear order in $\epsilon$. Indeed, after imposing (\ref{eq:intcond})
condition (\ref{eq:uvbrane}) is solved by
\begin{equation}
\text{Im} z = \epsilon D\theta D\bar{\theta} \left( 1 -\frac{\text{i}
    \theta \bar{\theta}}{2\epsilon D\theta D\bar{\theta}}\right) . 
\end{equation}
Plugging in (\ref{eq:tansatz}) yields the complete first order
correction
\begin{equation}
\text{Im} z = \epsilon \left( D\xi\right)^2 \left( 1 -\frac{
    \xi\rho}{\left( D\xi\right)^2}\right)+\ldots\, . 
\end{equation}
Finally, 
\begin{equation}
\rho =D^2\xi ,
\end{equation}
can be fixed by imposing (\ref{eq:intcond}). In summary, the solution
to the boundary conditions up to linear order in $\epsilon$ is
\begin{equation}
\theta = \xi + \text{i}\epsilon D^2\xi \,\,\, ,\,\,\, \text{Im} z =
\epsilon\left( 
\left( D\xi\right)^2  - \xi D^2\xi\right) .
\label{eq:expansion}
\end{equation}
The tangent vector in (\ref{eq:superex}) is computed as
\begin{equation}
T^1 = E^1_\xi \frac{\partial z^\xi}{\partial u} = e^\Sigma
\left(D\theta\right)^2 = \frac{1}{2\epsilon}
\frac{D\theta}{D\bar{\theta}} ,
\label{eq:tangentvec}
\end{equation}   
and $T^{\bar{1}}$ is obtained by complex
conjugation. Normalising\footnote{The non vanishing components of the
  2d flat metric and its inverse are $g_{1\bar{1}}=g_{\bar{1}1}=1/2$ and
  $g^{1\bar{1}}=g^{\bar{1}1} =2$.} 
$\left| n_1\right|^2 = n_1 n_{\bar{1}} =1/4$ and imposing $n_A T^A =0$
yields
\begin{equation}
n_1 = \frac{\text{i}}{2} \frac{D \bar{\theta}}{D\theta} .
\end{equation}
The first contribution to (\ref{eq:superex})  is
\begin{equation}
\frac{T^A D n_A}{\left| T^1\right|^2} =4\epsilon
\text{Im}\frac{D^2\theta}{D\theta}  = 4 \epsilon^2 \left( \frac{D^4
    \xi}{D\xi} - \frac{D^2 \xi D^3\xi}{\left( D\xi\right)^2}\right) ,
\end{equation}
where we suppressed higher orders in $\epsilon$. Next, we need to
compute the contribution due to the induced spin connection. To this
end, it is useful to notice that
\begin{equation}
\frac{\partial z^\xi}{\partial\vartheta}\Omega_\xi +
\vartheta\frac{\partial z^\xi}{\partial u}\Omega_\xi =
D\theta \left(\Omega_\theta +\theta \Omega_z\right) + D\bar{\theta}
\left(\Omega_{\bar{\theta}} +{\bar{\theta}} \Omega_{\bar{z}}\right)  ,
\end{equation}
The combination of spin connections is
exactly what is needed to make $D_\theta$ respectively
$D_{\bar{\theta}}$ covariant under local
rotations. In superconformal gauge this can be found in
\cite{Martinec:1983um}, 
\begin{equation}
\Omega_\theta +\theta \Omega_z = -D_\theta \Sigma
=D_{\bar{\theta}}\Sigma = \Omega_{\bar{\theta}} +\bar{\theta}
\Omega_{\bar{z}} =
\frac{\text{i}}{2\text{Im} z}\left( \bar{\theta} -\theta\right) .
\end{equation}
Plugging in expansion (\ref{eq:expansion}), we obtain (since normal
vector components are purely imaginary complex conjugation takes care
of the opposite charge under rotations) 
\begin{equation}
\left(\frac{\partial z^\xi}{\partial\vartheta}\Omega_\xi +
\vartheta\frac{\partial z^\xi}{\partial u}\Omega_\xi\right) \frac{T^1
  n_1}{\left| T^1\right|^2} +\text{c.c.}  = -4\epsilon^2   \frac{D^2
  \xi D^3\xi}{\left( D\xi\right)^2} .
\end{equation}
Thus we obtain for the extrinsic curvature in (\ref{eq:superex}) 
\begin{equation}
K = 4\epsilon^2 S\left[ t, \xi ; u,\vartheta\right] ,
\end{equation}
where the super-Schwarzian had been given in (\ref{eq:supersch}). This
is the expected result. We fixed the coefficient such that for $\eta
=0$ integration over 
$d\vartheta/2\epsilon^2$ (see (\ref{eq:epspow})) reproduces the
bosonic result. This 
seems reasonable 
since its bulk `partner' $\text{i}\int d^2\theta E R_{+-}$ gives just the bulk
part of the Gauss-Bonnet density in the bosonic case
\cite{Chamseddine:1991fg}.  

\section{Conclusions}

In the present note we considered Jackiw-Teitelboim supergravity. The
supercurvature is constrained to a constant value such that the actual
scalar curvature is constant and negative. The geometry is that of
$AdS_2$ superspace. We introduced a cutoff line near its boundary
by demanding that its arc length element differs from flat one
dimensional superspace only by a diverging constant factor. This
imposes conditions on superspace coordinates which resemble
superconformal transformations related to
super-reparameterisations of the boundary. The origins of the
respective conditions are quite similar due to 
demanding that the mixed fermion-boson component of the arc length has
a fixed ratio with the boson-boson component.
In addition, one finds that
the superspace boundary is given by a boson-fermion pair of functions
($f(u)$ and $\eta (u)$). We argued, that integrating over the bulk
leads to an effective super-Schwarzian Lagrangian for these
functions. Two arguments are given. First, we just assumed the
Lagrangian to be dilaton times super-Schwarzian. This could also be
viewed as an action for the boundary value of the dilaton. We checked
that the corresponding equations of motion hold when a bulk
solution of the dilaton is plugged in. The dilaton's boundary value is
obtained as the pull back to the boundary. This is consistent with
imposing Dirichlet conditions on variations and no further boundary
conditions. To ensure that Dirichlet conditions on variations
are enough to cancel boundary terms in variations, usually a
Gibbons-Hawking-York term is added. In our second argument for the
super-Schwarzian we supersymmetrised the bosonic Gibbons-Hawking-York
term. Indeed, when plugging in our solution for the boundary curve, it
gives rise to a super-Schwarzian Lagrangian. A divergence, present in
the bosonic case, is absent in the supersymmetric case. We gave
reasonable arguments for our normalisation of the supersymmetric
boundary term. It would be nice to confirm our term by direct
computation as carried out in Wess-Zumino gauge in
\cite{Grumiller:2009dx}. This as well as the extension to $N=2$
supersymmetry is left for future work. 

In summary, our note suggests the supergravity dual of the
supersymmetric SYK model to be given by nearly $AdS_2$
supergravity (containing the supersymmetric version of
Jackiw-Teitelboim gravity). In addition there will be massive
particles whose spectrum and interactions have been recently studied in
\cite{Gross:2017hcz} for the non supersymmetric SYK model. It would be
interesting to extend this to the supersymmetric case and see whether
supersymmetry leads to simplifications. 

\section*{Acknowledgements}  
This work was supported by the SFB-Transregio TR33 ``The Dark
Universe'' (Deutsche Forschungsgemeinschaft) and by ``Bonn-Cologne
Graduate School for Physics and Astronomy'' (BCGS).

\end{document}